\documentclass[aps,prl,preprint]{revtex4}

\usepackage{citesort}
\usepackage{graphicx}

\newcommand{\wn}{cm$^{-1}$}

\newcommand{\K}{\text{$\overline{\text{K}}$}\ }
\newcommand{\M}{\text{$\overline{\text{M}}$}\ }
\newcommand{\GM}{\text{$\overline{\Gamma\text{M}}$}\ }
\newcommand{\GK}{$\overline{\Gamma\text{K}}$}
\newcommand{\GKM}{\text{$\overline{\Gamma\text{KM}}$}\ }
\newcommand{\KM}{\text{$\overline{\text{KM}}$}\ }
\newcommand{\kpar}{\text{k$_\|$}\ }
\newcommand{\Angstr}{\text{\AA}}
\newcommand{\SO}{$S_{SO}$}

\begin{document}
%
%
\title{Adsorbate-induced Surface Stiffening: Surface Lattice
       Dynamics of Ru(001)-(1$\times$1)-O}

\author{T.~Moritz,$^{a}$ W.~Widdra\footnote{Corresponding author, Email:
widdra@e20.physik.tu-muenchen.de, Fax: +49-89-289-12338},$^{a}$
D.~Menzel,$^{a}$ K.-P.~Bohnen,$^{b}$ and R.~Heid$^{b}$}

\affiliation{$^{a}$Physik-Department E20, Technische
Universit\"{a}t M\"{u}nchen, D-85747 Garching, Germany}

\affiliation{$^{b}$Forschungszentrum Karlsruhe, Institut f\"ur
Festk\"orperphysik,  D-76021 Karlsruhe, Germany}

\date{December 21, 2000}

\begin{abstract}
The dynamical properties of the high-density
Ru(001)-(1$\times$1)-O phase has been investigated by a combined
high-resolution electron energy loss spectroscopy and density
functional theory study. Due to a strong static outward relaxation
of the first Ru layer a soft Rayleigh phonon mode is expected.
However, a Rayleigh mode stiffening together with a new high
energy in-plane phonon mode above the bulk bands is found which is
related to a strong adsorbate-induced intralayer force constant
stiffening which counteracts an interlayer softening. This
structurally rather simple system with one surface atom per
(1$\times$1) unit cell demonstrates the limited applicability of
previously adopted models.

\end{abstract}

\pacs{PACS~68.35Ja, 68.45Kg, 63.20Dj}

\maketitle

The dissociative chemisorption of oxygen on nickel, platinum and
ruthenium surfaces has been often treated as a model system for
reactive adsorbates on transition metal surfaces. Besides their
structural and electronic properties the {\it dynamical}
properties of oxygen adlayers have been the focus of several
studies, e.g. on Ni(100) \cite{Lehwald83,Szeftel84}, Pt(111)
\cite{Neuhaus87}, and Ru(001) \cite{He97,Moritz99,Kostov97}.
Within the various oxygen model systems, the (1$\times$1)-O
high-coverage phases play an important role due to their strong
influence on the properties of the first substrate layers, their
possible catalytic importance, and their unchanged adsorbate
overlayer periodicity \cite{Neuhaus87,Stampfl96}. Although they
are barely accessible under ultrahigh vacuum (UHV) conditions
using molecular oxygen, these dense adlayers have often been
claimed to bridge the gap in catalysis between studies under UHV
and ambient pressure/real conditions \cite{Stampfl96}.

For the Ru(001)-(1$\times$1)-O surface
just recently a strong oxygen-induced outward relaxation of 3.7~\%
for the first Ru layer was found based on a density functional
theory (DFT) and low energy electron diffraction (LEED) study
\cite{Stampfl96}. The (1$\times$1)-O induced outward relaxation is
in fact the strongest relaxation known on Ru(001)
\cite{Menzel99}. Such static structural changes which are driven
by a charge redistribution within the first layers will also
modify the dynamical properties of the surface. E.g., it is
commonly accepted that an outward relaxation of the first layer
will soften the Rayleigh wave
\cite{Lehwald83,Szeftel84,Berndt91,Rocca01}. Indeed, for the
Ru(001)-(1$\times$1)-O surface a peculiar soft vibrational mode
has been found and explained as a soft phonon mode due to the
first layer relaxation \cite{He97}. However, it could not be
confirmed in a subsequent high resolution electron energy loss
spectroscopy (HREELS) study \cite{Moritz99} questioning this
interpretation.

On the theoretical side the calculation of dynamical properties of
an adsorbate covered surface is rather demanding, because it
requires an accurate description of the adsorbate induced changes
of structural {\it and} electronic properties. Often, theoretical
descriptions of surface phonon spectra were based on empirical
force-constant models whose parameters were fitted to experimental
dispersion curves \cite{Rahman85,Berndt91,Braun97}. Recently, it
has been found that bulk Ru exhibits pronounced phonon anomalies
due to strong electron-phonon coupling \cite{Heid00}, indicating
the need for a large parameter set to accurately describe its
phonon spectrum. Previous models of the dynamics of clean and
adsorbate covered Ru(001) surfaces had used only a small number of
force constants \cite{Rahman83,Braun97} and are, therefore,
inadequate for a modeling of substrate dynamics. DFT based ab
initio methods provide an unbiased and often very accurate
description of vibrational properties. Early applications to
surfaces using frozen-phonon techniques were limited to phonons at
high-symmetry points of the surface Brillouin zone (SBZ)
\cite{Ho85} or to rough estimates of the interatomic coupling
\cite{Mueller86}. The way to a more complete calculation of
surface vibrations was opened by the development of numerically
efficient linear-response schemes in connection with
pseudopotential plane-wave techniques \cite{Baroni87,Giannozzi91}.
However, due to their localized $d$ electrons, transition metal
surfaces have been almost prohibitive to be dealt with by this
standard approach using a plane-wave basis only. These numerical
limitations have been overcome by a recent implementation of the
linear-response formalism in the framework of a mixed-basis
pseudopotential method \cite{Heid99}, where in addition to plane
waves also local functions are employed to better represent the
more localized parts of the valence states. This theoretical
technique which recently explained successfully the anomalous
lattice dynamics of bulk Ru \cite{Heid00}, is here applied to a
single crystal surface for the first time.

Clearly, the Ru(001)-(1$\times$1)-O surface is a benchmark for our
understanding of adsorbate-induced changes of the dynamical
surface properties. We present in this letter a combined,
theoretical DFT and experimental HREELS, study on the surface
phonon dispersion of the Ru(001)-(1$\times$1)-O surface. It will
be shown that the presence of an adsorbate layer induces a large
stiffening of the coupling within the first Ru layer, which gives
rise to a new substrate phonon appearing {\em above} the bulk
spectrum. This is in contrast to the Rayleigh mode which exhibits
only small changes due to a balancing of this intralayer
stiffening by interlayer softening.

\begin{figure}
\includegraphics[keepaspectratio=true,angle=0,width=0.98\textwidth]{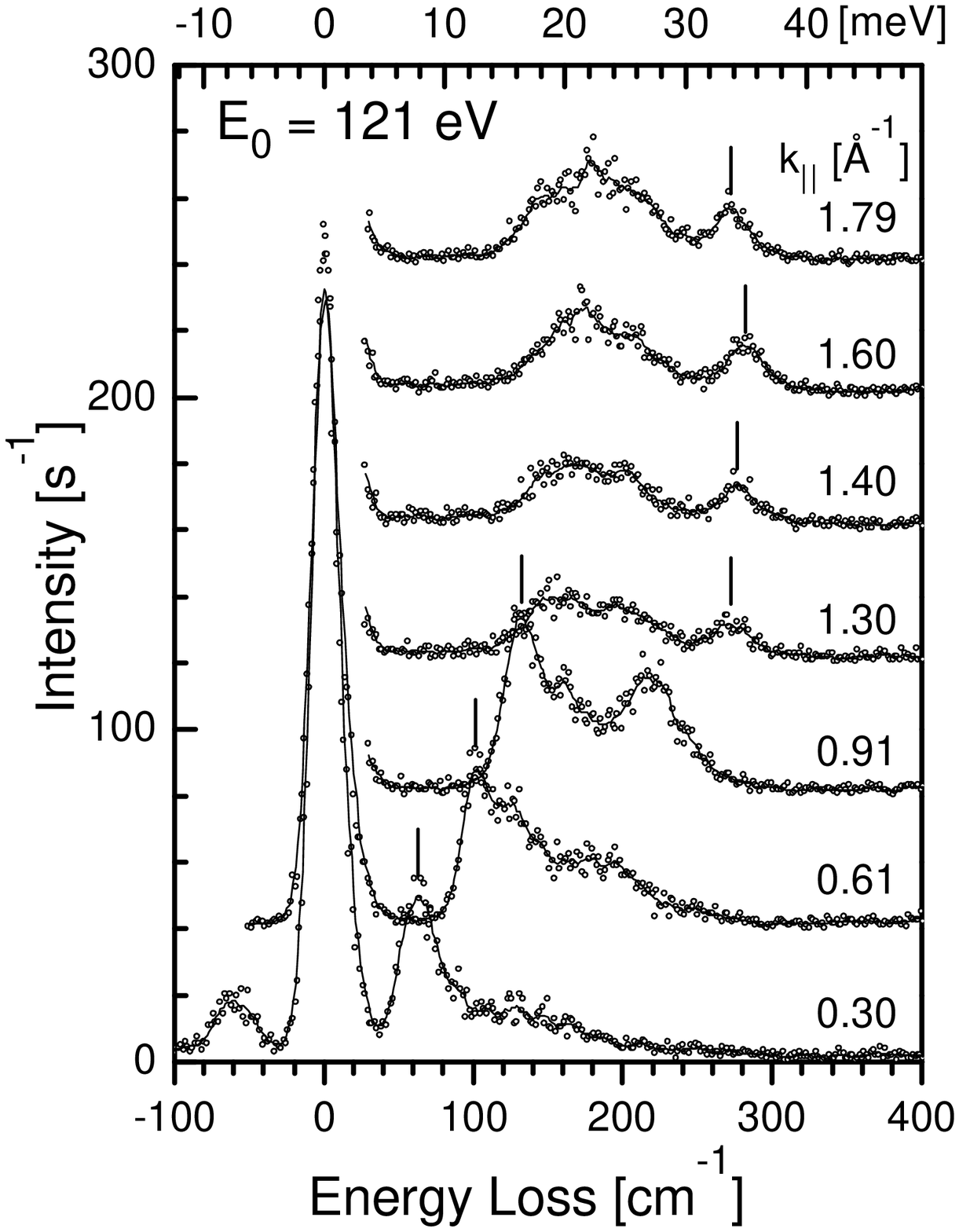}
\caption {HREEL spectra in the off-specular scattering geometry
for Ru(001)-(1$\times$1)-O at the indicated momentum transfer
along \GKM direction ([$10\overline{1}0$] direction). The electron
energy is 121~eV with a total scattering angle of 120$^{\circ}$.}
\label{hreelspectra}
\end{figure}

For the preparation of a well-ordered high-density O-(1x1) layer
dissociative adsorption of NO$_2$ at a surface temperature of
600~K has been used \cite{Stampfl96}. Preparation of this phase
with high exposures of molecular oxygen is also possible but leads
to less ordered  and incomplete (1$\times$1)-O layers due to the
necessity of vacant sites for the dissociative adsorption
\cite{Moritz99}. HREELS data for the Ru(001)-(1$\times$1)-O layer
in off-specular scattering geometry are shown in
Fig.~\ref{hreelspectra} for an electron energy of 121 eV and a
constant total scattering angle of 120$^\circ$. From bottom to top
the off-specular angle increases with the corresponding parallel
momentum transfer \kpar along the \GKM direction in reciprocal
space as indicated on the right side of Fig.~\ref{hreelspectra}.
Details of the experimental setup as well as a discussion of the
adsorbate modes above 500 \wn\ have been presented elsewhere
\cite{Moritz99}; here we focus on the substrate phonons. In
Fig.~\ref{hreelspectra} three prominent phonon modes can be easily
identified. At a momentum transfer of 0.3 \Angstr$^{-1}$ an energy
loss peak at 62~\wn\ dominates which is accompanied by a
corresponding energy gain peak at -62 \wn. This phonon mode shifts
up in energy to 101 and 130~\wn\ at 0.61 and 0.91 \Angstr$^{-1}$.
At higher momentum transfer this mode is difficult to identify at
the electron energy of 121~eV. In the momentum range from 1.3 to
1.79 \Angstr$^{-1}$ a new high energy mode around 280 \wn\ is
discernible. It initially shifts up in energy and then, for
momenta above 1.60 \Angstr$^{-1}$, down again. Additionally a
strong phonon loss at 230 \wn\ is visible at 0.91 \Angstr$^{-1}$.

\begin{figure}
\includegraphics[keepaspectratio=true,angle=90,width=0.98\textwidth]{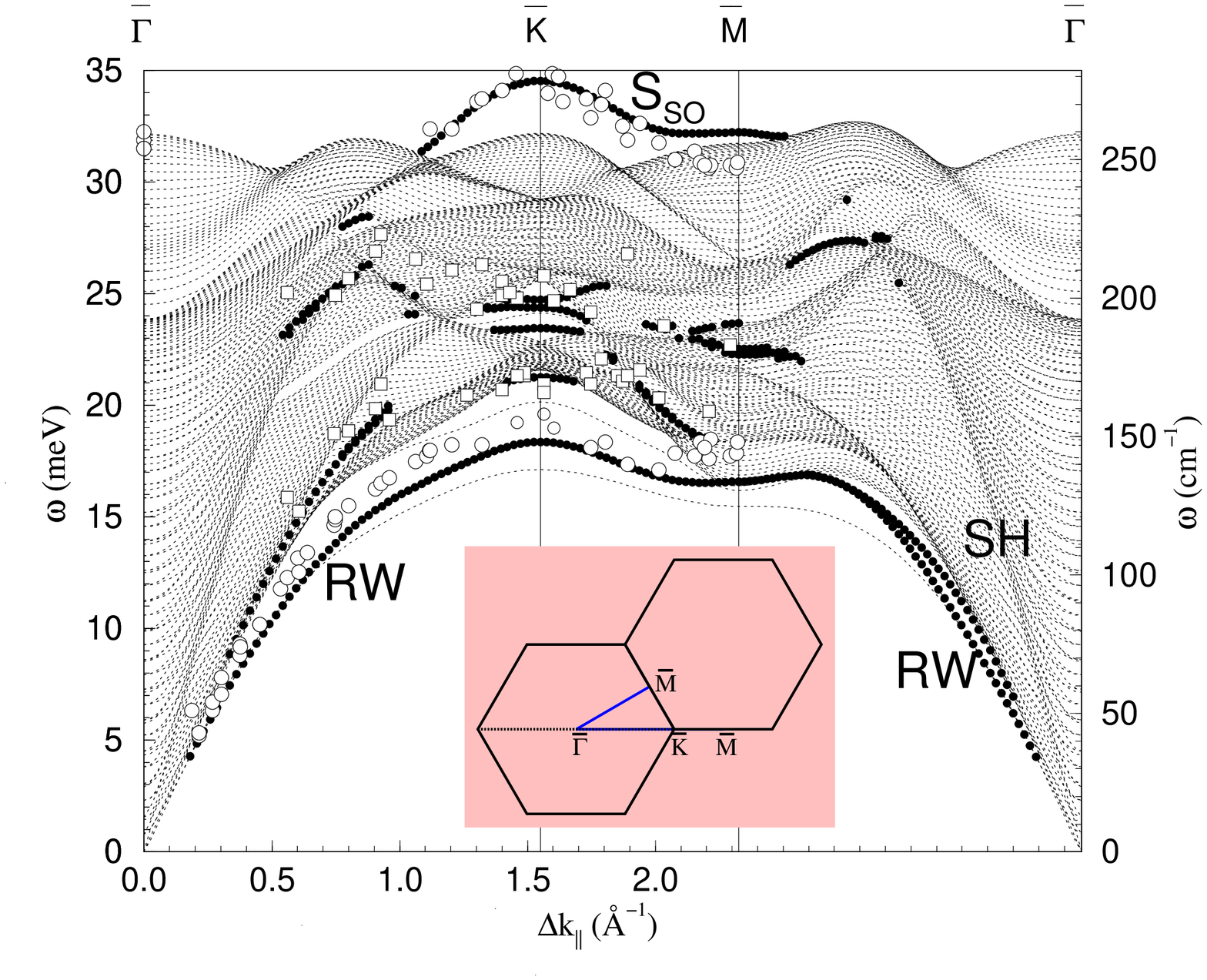}
\caption {Phonon dispersion curves for the Ru(001)-(1$\times$1)-O
surface. Open symbols indicate the experimental data; theoretical
dispersion curves for an asymmetric 50 layer slab are indicated by
dashed lines, while filled circles represent surface modes (identified
by a weight larger than 20 \% in the first three layers).
In the inset the hexagonal SBZ is sketched.} \label{dispersion}
\end{figure}

The experimental phonon frequencies extracted using various
electron energies in the range of 16 to 256~eV and various \kpar
are summarized by open symbols in Fig.~\ref{dispersion}. The
(1$\times$1) SBZ is shown schematically in the inset. The dashed
lines and the solid circles in Fig.~\ref{dispersion} indicate the
phonon dispersion curves for the surface projected bulk and the
surface phonons, respectively, as determined by first principles
linear-response calculations. The theoretical results correspond
to an asymmetric 50 layer slab with an adlayer covered surface on
one side and a bulk truncated surface on the opposite side of the
slab. Ab initio force constants for Ru bulk taken from
Ref.~\cite{Heid00} have been combined with surface force
constants, which were extracted from mixed-basis linear-response
calculations \footnote{The linear-response implementation from
\cite{Heid99} is based on the
  "Fortran90 Program for
  Mixed-Basis Pseudopotential Calculations for Crystals" by
  B. Meyer, C. Els\"{a}sser, and M. F\"{a}hnle,
  Max-Planck-Institut f\"{u}r Metallforschung, Stuttgart (unpublished).}
for a symmetric slab consisting of 6 Ru layers and one O layer
added on each side. All but the two innermost layers have been
relaxed. The local-density approximation using the Hedin-Lundqvist
form of the exchange-correlation functional has been applied
\footnote{Details of the norm-conserving pseudopotentials and
local functions have been described in previous works
\cite{Heid00,Heid00b}. A cutoff of 22 Ry has been employed for the
plane-wave basis, and the SBZ sampling was performed on a 18x18
hexagonal $k$-point mesh corresponding to 37 $k$ points in the
irreducible wedge.}.

In the calculations several modes (marked with solid circles) show
strong localization at the surface, i.e. exhibit high amplitude
weights within the first Ru layers.
The most striking feature is the high energy split-off surface
phonon (\SO) which is peeled off from the bulk bands around the \K
point and which is located at approximately 280~\wn\ (35~meV)
there.
At the \K point it is located 19~\wn\ higher than any bulk phonon
mode. It is important to note that calculations based on the force
constants of {\it bulk} Ru and the geometrical structure of the
truncated crystal do not show any surface phonon located above the
bulk bands. Similarly previous theoretical studies of the hcp
(001) or fcc (111) surface phonons did not find such a mode
\cite{Allen72,Allen71b}. Therefore this split-off phonon indicates
significantly changed dynamical properties of the (1$\times$1)-O
surface in comparison to the Ru bulk.

The second feature which is experimentally clearly visible is
attributed to the Rayleigh wave (RW).
The experimentally observed phonon dispersion for \kpar $>$ 0.2
\Angstr$^{-1}$ agrees within 10~\wn\ with the calculated
dispersion curve. The comparison with the bulk phonon bands shows
that it is located well below the bulk bands as observed for many
other surfaces \cite{Rocca01}. Note however that the dispersion of
the RW is shifted upwards in energy (for \kpar $>$ 0.7
\Angstr$^{-1}$) as compared to the results for the truncated bulk
in the absence of oxygen (lowest dashed line).
Besides these two modes which are visible within a wider part of
the SBZ, modes at 170 and 200~\wn\ (21 and 25~meV) are
experimentally discernible around the \K point which can also be
found in the calculated surface phonon dispersion curves.

\begin{figure}
\includegraphics[keepaspectratio=true,angle=0,width=0.98\textwidth]{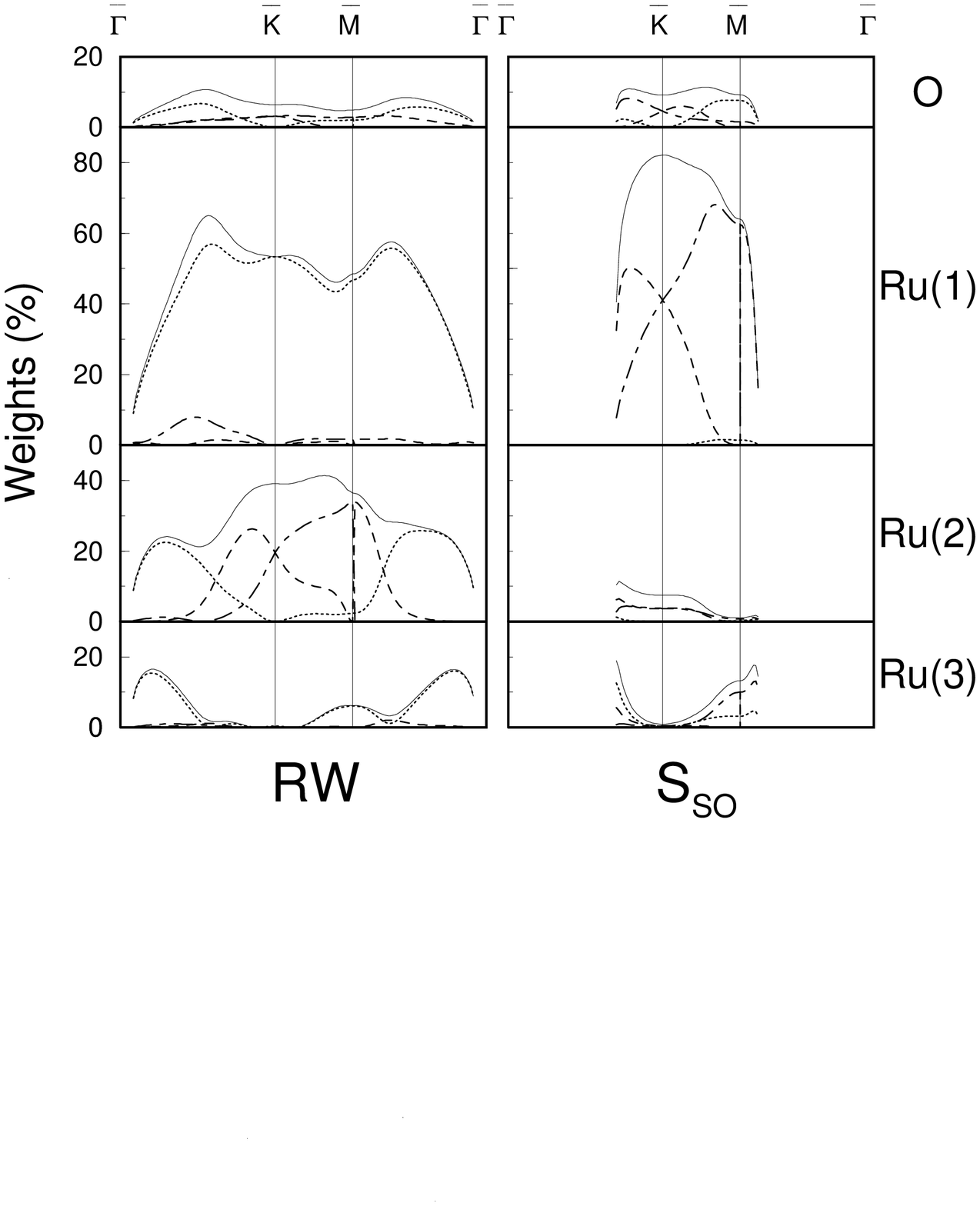}
\caption {Relative layer-resolved vibrational weights for the
Rayleigh (left panel) and the high energy split-off mode (right
panel) along \GK, \KM, and \GM. Shown are the total sum (full
lines) and the weights of the vertical, longitudinal and shear
horizontal components (dotted, dashed, and dash-dotted lines,
respectively). } \label{weight}
\end{figure}

For the RW phonon and the \SO\ mode, the symmetry- and
layer-resolved vibrational amplitudes are displayed in
Fig.~\ref{weight}. From its dominantly vertical character in the
first Ru layer throughout the whole SBZ the Rayleigh mode (left
side of Fig.~\ref{weight}) has been identified. It corresponds to
the prominent $S_1$ mode which has been described for fcc(111) and
hcp(001) surfaces \cite{Allen71,Allen72,Black83}. Along \GK,
admixtures of longitudinal and shear horizontal character can be
seen which are small for the first layer but substantially for the
second layer Ru atoms. This mixing is possible because the \GK\
direction is a low symmetry (C$_1$) direction. On the other hand,
along the \GM direction the Rayleigh mode
involves only atomic motions within the sagital plane (vertical-longitudinal)
because this direction lies within
a surface mirror plane (has C$_S$ symmetry). At the \K point which
exhibits C$_{3V}$ symmetry, the movement of the O and Ru atoms in
each layer are either purely vertical or purely in plane. For the
Rayleigh mode we find vertical movements of the first layer Ru
atoms and in-plane movements for the O and the second layer Ru
atoms. It is strongly localized within the first and second Ru
layers here. The mode which is visible around the \K point at
170~\wn\ (just below the bulk bands) corresponds to the $S_5$ mode
of Ref. \cite{Allen72} and has a similar displacement pattern; but
here the first layer Ru atoms move in plane and the O and second
layer Ru atoms move vertically. As for the RW the S$_5$ phonon is
shifted upwards in energy as compared to the truncated bulk
calculations (dashed lines in Fig.~\ref{dispersion}).

For the high energy mode at the \K point (right side of
Fig.~\ref{weight}), the movements of the O and Ru atoms are
completely in plane circular for all layers. When the O atom moves
counter clockwise, the first layer Ru atoms move clockwise, or
vice versa. For the deeper Ru layers the pattern repeats. The
movement is mainly located in the first Ru layer with only minor
contributions from second or deeper layers. However it has also
appreciable amplitude on the O atoms despite the large Ru to O
mass mismatch. This indicates some hybridization with the Ru-O
bending modes although the adsorbate modes are a factor of two
higher in frequency (520 to 565~\wn). Close to the \M point the
phonon mode exhibits some vertical movement of the O atoms and a
dominantely shear horizontal movement of the first layer Ru atoms
which renders the experimental identification of this mode
difficult close to the \M point.
Because of the pure in-plane movement of this mode its high energy
at \K points to a significant stiffening of the {\it intralayer}
force constants due to the oxygen overlayer. This is indeed
confirmed by the detailed theoretical analysis of the interatomic
force constants. They exhibit a large increase of the nearest
neighbor coupling within the first Ru layer by a factor of 2.5
compared to the bulk.

The adsorbate-induced changes of the surface dynamical properties
can be related at least partly to its changed geometrical and
electronic structure: For the Ru(001)-(1$\times$1)-O surface we
find a large outward relaxation (3.83~\%) of the first-to-second
Ru layer spacing, d$_{12}$, with respect to the bulk layer spacing
which is accompanied by a contraction of the second-to-third layer
spacing of -0.47~\%. This finding is in perfect agreement with a
previous DFT-GGA calculation and recent LEED studies which found
values of 2.7 and 3.7~\% for d$_{12}$ and -0.9 and -0.5~\% for
d$_{23}$, respectively \cite{Stampfl96,Menzel99}. It has been
shown that the large expansion of the first Ru interlayer spacing
is the result of a weakening of the attraction between the top and
second substrate layers due to an oxygen-induced charge removal
from the Ru {\it d} states \cite{Stampfl96}. Consistent with this
picture, we find from the theoretical analysis that the outward
relaxation indeed weakens the {\it interlayer} force constants
between the first and second Ru layers which alone would imply a
RW softening. Such a softening has been found for several
examples, e.g. for oxygen on Ni(100), Cu(100), and
Pd(100)~\cite{Szeftel84,Rahman85,Chen91,Chen94}. However we find
here -- against this generally assumed rule \cite{Rocca01}-- a
Rayleigh phonon {\it stiffening} in the presence of a strong
outward relaxation. The explanation is closely related to the
observation of the new \SO\ mode. The intralayer stiffening which
at the \K point peels off the \SO\ mode from the bulk bands
counteracts the softening of the RW. It even leads to an
over-compensation near the SBZ boundary where the intralayer
couplings are more effective than in the limit of long
wavelengths. Thus the unexpected stiffening of the RW phonon and
the \SO\ mode around the \K point are both consequences of the
adsorbate-induced intralayer stiffening.
Part of this stiffening might be of electrostatic origin. It is
known from theoretical studies of O adsorbates on Rh
\cite{Stokbro97,Ganduglia99}, Pt \cite{Feibelman97,Kokalj99}, and
Ru surfaces \cite{Feibelman99} that due to the large
electronegativity of oxygen the O-metal bond is slightly ionic,
i.e.\ there exists a small charge transfer from the metal surface
to the O layer. For example, a charge transfer of 0.3 electrons
per surface atom has been proposed for Pt(111)-(1$\times$1)-O
\cite{Kokalj99}. Consequently, the substrate is somewhat
positively charged leading to an enhanced Coulomb interaction.
Indeed, we find that for the nearest neighbor bond of the first Ru
layer, the changes in the diagonal components of the force
constant matrix (longitudinal and transverse force constants) with
respect to the bulk have the form of a Coulomb mediated coupling.
However, this explanation would require effective charges of more
than one electron per Ru site, which seems to be unreasonably
large. In addition, we also find significant changes in the
off-diagonal (tensorial) components of the force constant matrix.
Thus, the adsorbate induced modifications of the binding
properties are rather complex and involve many-body interactions
as well.

To summarize, the different behaviour of intra- and interlayer
force constants for the Ru(001)-(1$\times$1)-O surface shows
clearly that an adsorbate covered surface -- even in the simplest
case of unchanged symmetry and only a single atom per layer in the
unit cell -- cannot be described with force constants which are
scaled from the respective bulk values. Here the dense adlayer
introduces a strong intralayer force constant stiffening in
combination with an interlayer softening. It has been demonstrated
that the former leads to a new high energy in-plane phonon mode
above the bulk bands. For the RW mode the intralayer stiffening
counteracts the phonon softening which might be expected otherwise
on the simple basis of first-layer relaxation.

This work has been supported by the Deutsche
Forschungsgemeinschaft through SFB 338.


\clearpage

%

\end{document}